\documentclass[10pt,conference]{IEEEtran}
\usepackage{cite}
\IEEEoverridecommandlockouts
\usepackage{tikz}
\usepackage{amsmath,amssymb,amsfonts}
\usepackage{algorithmic}
\usepackage{graphicx}
\usepackage{caption} 
\usepackage{textcomp}
\usepackage{xcolor}
\usepackage{soul}
\usepackage{xspace}
\usepackage{url}
\usepackage{booktabs}
\usepackage{enumitem}
\usepackage{tcolorbox}
\usepackage{colortbl} 
\usepackage{hyperref}
\usepackage{cleveref}
\usepackage{listings} 
\usepackage{graphicx}
\usepackage{subcaption}
\def\BibTeX{{\rm B\kern-.05em{\sc i\kern-.025em b}\kern-.08em
    T\kern-.1667em\lower.7ex\hbox{E}\kern-.125emX}}

\definecolor{ghostwhite}{rgb}{0.97, 0.97, 1.0} 
\usepackage{tikz}
\newcommand{\circled}[1]{\tikz[baseline=(char.base)]{
  \node[draw, circle, inner sep=1.1pt, line width=0.5pt]
  (char) {\sffamily\bfseries\footnotesize#1 };}}

\newcommand{\para}[1]{%
  \noindent\textbf{#1.\ }%
}

\crefformat{section}{§#2#1#3}
\Crefformat{section}{§#2#1#3}

\newcommand{\gin}{\textit{Gin}\xspace}
\newcommand{\diffP}{\textit{patchDiff}\xspace}

\newcommand{\model}{\textit{PatchCat}\xspace} 
\newcommand{\sumLLMshort}{\textit{briefSum}\xspace}
\newcommand{\online}{application\xspace}
\newcommand{\offline}{training\xspace}

\newcommand{\Mistral}{\texttt{Mistral}\xspace}

\newcommand{\jcodec}{\texttt{JCodec}\xspace}
\newcommand{\commonsnet}{\texttt{Commons-Net}\xspace}

\newcommand{\junit}{\texttt{JUnit4}\xspace}

\newcommand{\manTagged}{309\xspace}
\newcommand{\numRaw}{11{,}381\xspace}
\newcommand{\numUnique}{5{,}806\xspace}
\newcommand{\sampleVal}{218\xspace}

\newcommand*{\eg}{e.g.,\@\xspace} 
\newcommand*{\ie}{i.e.,\@\xspace}

\usepackage[textwidth=\linewidth]{todonotes}
\usepackage{xcolor}

\usepackage{titlesec}
\titlespacing*{\section}{0pt}{0.1em}{0.1em}
\titlespacing*{\subsection} {0pt}{0.1em}{0.1em}
\titlespacing*{\subsubsection} {0pt}{0.1em}{0.1em}
\begin{document}

\title{LLM-Guided Genetic Improvement: Envisioning Semantic Aware Automated Software Evolution}

\author{
\IEEEauthorblockN{Karine Even-Mendoza}
\IEEEauthorblockA{\textit{King's College London}, UK \\
}
\and
\IEEEauthorblockN{Alexander Brownlee}
\IEEEauthorblockA{\textit{University of Stirling}, UK \\
}
\and
\IEEEauthorblockN{Alina Geiger}
\IEEEauthorblockA{\textit{JGU Mainz}, Germany \\
}
\and
\IEEEauthorblockN{Carol Hanna}
\IEEEauthorblockA{\textit{University College London}, UK \\
}
\and
\IEEEauthorblockN{Justyna Petke}
\IEEEauthorblockA{\textit{University College London}, UK \\
}
\and
\IEEEauthorblockN{Federica Sarro}
\IEEEauthorblockA{\textit{University College London}, UK \\
}
\and
\IEEEauthorblockN{Dominik Sobania}
\IEEEauthorblockA{\textit{JGU Mainz}, Germany \\
}
}

\maketitle

\begin{abstract}

Genetic Improvement (GI) of software automatically creates alternative software versions that are improved according to certain properties of interests (e.g., running-time). 
Search-based GI excels at navigating large program spaces, but operates primarily at the syntactic level. In contrast, Large Language Models (LLMs) offer semantic-aware edits, yet lack goal-directed feedback and control (which is instead a strength of GI).

As such, we propose the investigation of a new research line on AI-powered GI aimed at incorporating semantic aware search.
We take a first step at it by augmenting GI with the use of automated clustering of LLM edits. 

We provide initial empirical evidence that our proposal, dubbed \model{}, allows us to automatically and effectively categorize LLM-suggested patches. \model{} identified 18 different types of software patches and categorized newly suggested patches with high accuracy. It also enabled detecting \textit{NoOp} edits in advance and, prospectively, to skip test suite execution to save resources in many cases.  These results, coupled with the fact that \model{} works with small, local LLMs, are a promising step toward interpretable, efficient, and green GI. 

We outline a rich agenda of future work and call for the community to join our vision of building a principled understanding of LLM-driven mutations, guiding the GI search process with semantic signals.

\end{abstract}


\section{Introduction}

Large language models (LLMs) are reshaping automated software improvement by enabling the generation of high-level, semantically rich program edits \cite{LLMSE:icse:2023:survey,ssbse2023evaluating,DBLP:conf/ssbse/BrownleeCEGHPSS23,10.1109/TEVC.2024.3506731,wang2025large,DBLP:journals/ase/BrownleeCEGHPSS25}. 

In Genetic Improvement (GI), this opens a powerful new source of mutations that go beyond syntax-level rewrites and fixed mutation rules, thus enabling more flexible exploration \cite{wang2025large,DBLP:journals/ase/BrownleeCEGHPSS25}. 
However, integrating LLMs as a mutation operator in GI introduces a new challenge: Mutations are often noisy, redundant, or unrelated to the performance goal, commonly leading to \textit{NoOp edits} (no observable behavioral change) or \textit{invalid patch code} \cite{Fuzz4All2023,10.1145/3691620.3695529}. 
Evaluating each LLM-generated patch through full compilation and testing is expensive and wasteful, and repeatedly generating redundant or similar patches leads to inefficiency and environmental waste \cite{campos2025empiricalevaluationgeneralizableautomated,petke2023program,8753481}.
Hence, while LLMs expand the search space, the process lacks effective guidance to prioritize noteworthy patches.
The underlying dynamics of how and why LLM-generated patches work remain poorly understood. We lack a systematic and efficient way to characterize the types of edits produced, how they differ from traditional mutations, and which are most effective in different contexts.
Such an understanding is a crucial step towards building solutions in this new AI-powered genetic improvement research agenda: ones that combine LLM creativity with guidance and understanding.

In this paper, we propose a new viewpoint for LLM-guided GI to bridge the gap: The use of a lightweight machine learning model for fast approximation of patch edits as a means to interpret the semantic nature of patches (\eg{} ``added exception handling", ``renamed variable") and guide mutation decisions during search. 
This model acts as a fast filter, prioritizing promising edits and discarding low-value candidates before costly evaluation, and moving away from blindly sampling patch edits from an LLM’s output.

Building on previous work using Sentence-BERT and short-text clustering~\cite{reimers2019sentence,10.1007/978-3-030-51310-8_10}, we introduce \model{} (see \cref{approach}), trained on manually tagged LLM-generated GI patches of open source projects in Java \cite{DBLP:journals/ase/BrownleeCEGHPSS25} to group patch summaries. Using \model, we characterize the nature of these edits and aim to identify patterns in successful mutations and map the space of effective LLM-driven transformations.

We provide initial experimental results (see (\cref{results})) showing the ability of \model{} to successfully predict \textit{NoOp edits} within the 18 discovered categories of patch edit types, derived from an initial 23 through manual tagging. In this process, two categories were merged and four were excluded for insufficient data.
We examined their frequency and quality, with our prototype achieving $\sim$66\% accuracy on a validation set of unseen data, which can be further improved with additional steps of data cleaning, category expansion, and enhanced supervision.

This exploratory analysis sets the stage for future work on using structural insights from patch clustering to inform mutation filtering, guide search more effectively, or even steer LLMs toward generating higher-quality patches from the outset.

In addition to this empirical evidence, we present a comprehensive research agenda (see \cref{researchAgenda}) toward the ultimate goal of AI-powered genetic improvement.

\model, the intermediate trained models, code and training scripts, and datasets used in this study are available online as an open-access artifact \cite{anon_2025_15834984}.

\clearpage

\section{Semantic Classifier for LLM-Guided GI}
\label{approach}

We evaluate the quality of LLM-generated patches~\cite{DBLP:conf/ssbse/BrownleeCEGHPSS23,DBLP:journals/ase/BrownleeCEGHPSS25}, produced via \gin{}~\cite{brownlee2019gin}, a tool designed to facilitate experimentation with GI techniques by automating the process of transforming, building, and testing Java projects. The dataset of patches was generated using an LLM-based mutator with three LLMs on two large-scale projects~\cite{brownlee_2024_projects_dataset}.

\para{Training Phase}
(\cref{sec:modelextraction} \& Fig.~\ref{fig:offline}) 
To analyse these patches, we perform a multi-step process that characterizes their semantic nature and evaluates their impact. Based on this analysis, we build our Automated Patch Classification model, \model{}, to categorize patches into meaningful clusters, representing different types of edits (\eg{} control flow changes, comment modifications, API replacements). 

\para{Application}
(\cref{sec:auto} \& Fig.~\ref{fig:online})
Using \model{}, we suggest augmenting the \gin{} search process to improve the efficiency of patch evaluation. We implemented \model{} as a light-fast approximation model of the patch's semantics in the \offline phase, and discuss its integration in the research agenda (\cref{researchAgenda}).\looseness=-1

\subsection{Patch Classification}
\label{sec:modelextraction}

We outline the creation of \model{} as the \offline{} phase in Fig.~\ref {fig:offline}. 
To train \model{}, we curated a dataset that contains the \textit{description of a patch} (meaning the difference between the original code version and the patched code) and the category to which we assign the patch.
For brevity, we denote the description of a patch as \diffP{}.

\para{Generate Patch Data}%
    (Fig.~\ref{fig:offline}, \circled{A})  
    We generate \diffP{}  for each patch suggested by \gin{} using \texttt{diff original\_i.java patched\_i.java}.

\para{Manually Describe Patches}%
    (Fig.~\ref{fig:offline}, creating \circled{B}) 
    We manually annotate each \diffP{} with a \textit{short natural language description}, capturing the essence of the change, producing a paired dataset of \diffP{} and \sumLLMshort{}.

\para{Manually Categorize Description}%
    (Fig.~\ref{fig:offline}, creating \circled{C})
    We manually assign descriptions to categories (\eg{} comment deletion, control flow change, API replacement), creating a labeled dataset that contains common types of patch intents. 
    This produces a paired dataset of 
    \textit{Category ID} and \sumLLMshort{}. 
    
\noindent
    We manually tagged \manTagged{} entries,  
    limited to the \Mistral generated patches, and \junit{} and \jcodec{} projects (those with the highest pass rates in  
    Brownlee et al.~\cite{DBLP:journals/ase/BrownleeCEGHPSS25,brownlee_2024_13381774_zenodo}'s work).

\para{Data Augmentation}%
(Generate more patch descriptions per category, Fig.~\ref{fig:offline},  creating \circled{D}) 
For each category, we wrote a small Python program to generate synthetic summaries. This enriches the dataset with labeled, synthetic training data. We manually reviewed the results to confirm they were sensible. 
This yields a dataset of \numRaw{} entries, which is reduced to \numUnique{} unique entries used to train \model.

\para{Building \model{}}%
(Train model with descriptions and categories, Fig.~\ref{fig:offline},  creating \circled{E}) 
We used the dataset from the previous phase to train \model{}. 
After splitting the data of \numUnique{} unique entries into training and test sets, we applied a semi-supervised clustering approach~\cite{10.1007/978-3-030-51310-8_10}, previously shown to improve short text clustering on StackOverflow question titles~\cite{XU201722}.
In this work, we adapt the approach in~\cite{10.1007/978-3-030-51310-8_10} to short descriptions of code patches. We use Python's K-Means \cite{sklearn-kmeans} as a baseline.  
We embed all summaries using Sentence-BERT (MiniLM) \cite{reimers2019sentence,NEURIPS2020_3f5ee243}.
We measured accuracy at 0.792 and Normalised Mutual Information (NMI) at 0.735 of the baseline. After enhancing with \cite{10.1007/978-3-030-51310-8_10}, the cluster alignment was with accuracy = 0.787, NMI = 0.741. The trained model and vectorizer form \model{}.

\subsection{Automation}%
\label{sec:auto}

\begin{figure*}[t]
\centering

\begin{minipage}[t]{0.68\textwidth}
  \vspace{0pt} 
  \begin{subfigure}[t]{\textwidth}
    \centering
    \includegraphics[width=1.03\textwidth,trim=5pt 2pt 2pt 8pt]{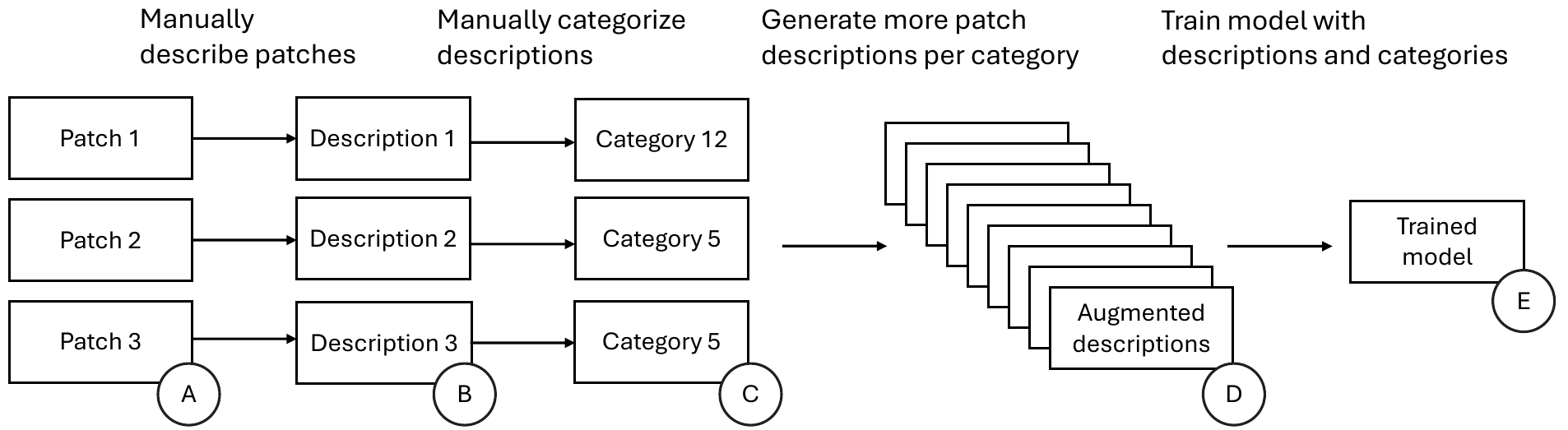}
    \vspace{-0.5cm}
    \caption{Training}
    \label{fig:offline}
  \end{subfigure}
  \vspace{0.2cm}
  \begin{subfigure}[t]{\textwidth}
    \centering
    \includegraphics[width=0.99\textwidth,trim=5pt 8pt 2pt 1pt]{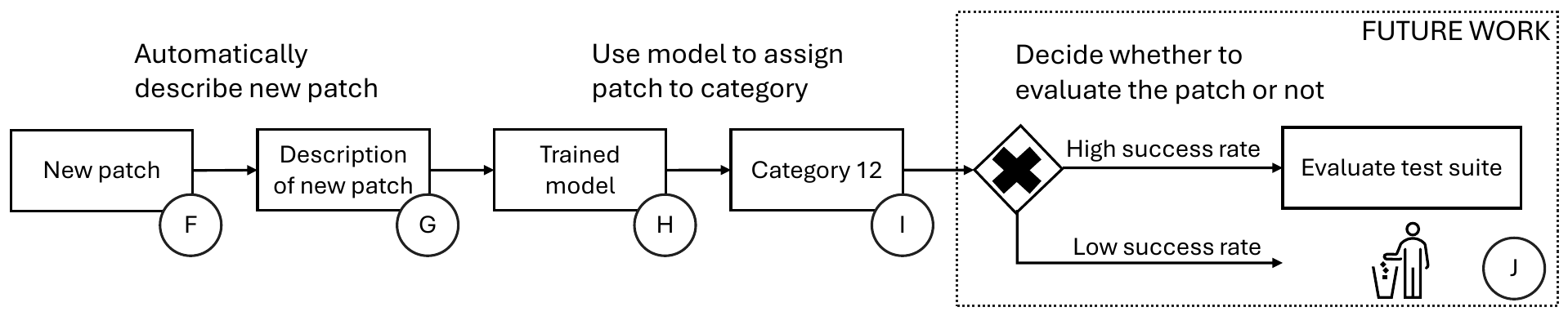}
    \vspace{-0.6cm}
    \caption{Application}
    \label{fig:online}
  \end{subfigure}
  \caption{
  Training (a) and Application (b) Scenarios.
    \footnotesize
    In the \offline phase, each patch is manually described and categorized. These descriptions are used to generate augmented data to train \model{}.
    In the \online phase, \model{} automatically assigns patches to categories. Only those in promising categories are tested; the rest are discarded.}
\end{minipage}%
\hfill
\begin{minipage}[t]{0.31\textwidth}
  \vspace{0pt} 
    \centering
       \captionof{table}{Category (\#) \& Description.}
    \scriptsize
\begin{tabular}{cp{0.82\textwidth}}
\toprule
\# & 
Description \\
\midrule
0 & Added (some arbitrary) code from GitHub \\
1 & No change \\
2 & Modified a comment (add/remove/edit) \\
3 & Deleted blocks in a method (all/most/some) \\
4 & Duplicate code \\
5 & Modifications to return statements (add/remove/edit)  \\
6 & Changes to method names  \\
7 & Changed data types or type usage and generics  \\
8 & Includes inlining of implementations (sort, methods...)  \\
9 & Added exception-handling constructs (unreachable or reachable)\\
10 & Added extra brackets  \\
11 & Added synchronization logic \\
12 & Modified Variable/Class/Object Name  \\
13 & Modified Control Flow Structure \\
14 & Modified Object/Primitive Creation or Initialization  \\
15 & Split a statement into multiple lines  \\
16 & Arithmetic manipulation (boolean var. or expr. manipulations) \\
17 & Added dead code  \\

\bottomrule
\end{tabular}
    \label{tab:offline:res:15words}
\end{minipage}
\vspace{-0.5 cm}
\end{figure*}

We outline the usage of \model{} in \gin{} as the \online phase in Fig.~\ref {fig:online}. 
The \online phase aids the automatic analysis of patches post-execution of \gin (as discussed in RQ2 in~\cref{results}).
This phase enables the collection of statistics on patch quality and diversity of generated patches across large codebases. 
In~\cref{researchAgenda}, we discuss extending this approach to not only automate post-processing but further integrate it as a feedback-guided component within \gin{}. 

\para{New Patch}%
(Fig.~\ref{fig:online}, \circled{F}) 
The process begins with a patch produced by \gin{}. We create a \diffP{} with the original and the \gin-mutated patched code, \eg{} Listing \ref{lst:diff-1}.
\begin{lstlisting}[backgroundcolor=\color{ghostwhite}, basicstyle=\tiny,
caption={\diffP{} of a patch generated with \Mistral for the project \commonsnet{}.}, label={lst:diff-1}, captionpos=b, aboveskip=5pt, belowskip=2pt]
312c312,313
<             throw new ParseException(``Timestamp '' + timestampStr + `` could not be parsed using a server time of '' + serverTime.getTime().toString(), pp.getErrorIndex());
---
>             String errorMessage = ``Timestamp '' + timestampStr + `` could not be parsed using a server time of '' + serverTime.getTime().toString();
>             throw new ParseException(errorMessage, pp.getErrorIndex());
\end{lstlisting}

\para{Automatically Describing a New Patch}%
(Fig.~\ref{fig:online}, creating \circled{G})
To facilitate further automation, we implemented a script that, given a \diffP{} 
generates a short summary of the changes, \ie{} \sumLLMshort{}, using the \texttt{llama3} LLM and the prompt in Listing \ref{lst:prompt-1}. This summary abstracts away irrelevant syntactic changes and highlights the core modification.
\begin{lstlisting}[backgroundcolor=\color{ghostwhite}, basicstyle=\footnotesize,
caption={Prompt to generate a \sumLLMshort{} of \diffP{}}, label={lst:prompt-1}, captionpos=b, aboveskip=5pt, belowskip=2pt]
``Summarize the following Java diff in exactly 15 words: <<diff>>"
\end{lstlisting}
 
\para{Using \model to Assign Patch to a Category}
(Fig.~\ref{fig:online}, 
receiving as input \circled{H}, 
invoking \circled{I} and 
outputting \circled{J})
The \sumLLMshort{} is then passed to \model, which maps it to a predefined semantic category (\eg{} \textit{``just added try and catch''} to Category 9 as exemplified in Listing~\ref{lst:model-usage-example}). 
\begin{lstlisting}[backgroundcolor=\color{ghostwhite}, basicstyle=\footnotesize,
caption={Examples of querying \model{}.}, 
label={lst:model-usage-example}, captionpos=b, aboveskip=5pt, belowskip=2pt]
$ ... ``nothing much there is no difference really" . . => [1]
$ ... ``seems like there are only new comments" . . . => [2]
$ ... ``just added try and catch" . . . . . . . . . . . . . . .=> [9]
\end{lstlisting}

\para{\model-Guidance Enhancing AI-powered GI}
(FUTURE WORK dashed box, Fig.~\ref{fig:online}, \circled{J} as use case example) 
This classification could be used in the future to inform downstream decisions.

\para{Example A: NoOp Patches}
\textit{No Operation} (\textit{NoOp}) patches do not change the behavior of the program or are dead code. Hence, patches of Categories 1, 2 and 17 (see Table ~\ref{tab:offline:res:15words}) are an example of NoOp.
In practice, we would like to keep only the patches that compile and pass tests that are not simple \textit{NoOps}.

\noindent%
As the patch interpretation in \gin{} will be part of our future work, we will discuss this further along with more complex scenarios in the research agenda in~\cref{researchAgenda}.

\section{Preliminary Evaluation}
\label{results}
Our classified patches are taken from Brownlee et al.~\cite{DBLP:journals/ase/BrownleeCEGHPSS25,brownlee_2024_13381774_zenodo}'s work, which recorded whether each patch could be \textit{compiled}, whether the resulting patched code still \textit{passed} all unit tests, and whether it was in fact a \textit{NoOp}.

In this work, we evaluate two core aspects: the accuracy of \model and the nature of LLM-based patches in \gin, each tied to one research question (RQ).
Therefore, we ask:
\begin{list}{}{\leftmargin=0.7em \rightmargin=0.9em}
\item
\textbf{RQ1:} \textit{How accurately does the model assign categories to patches based on the \diffP{} between original and patched code, and how reliably can these categories inform decision-making in \gin{}?}

\item
\textbf{RQ2:} \textit{What types of patches does \gin{} attempt to apply when mutating code via LLM-generated edits, and what is their quality in terms of diversity, static validity (\ie{} compilation rate), and dynamic validity (\ie{} passing the software under optimisation's test suite)?}
\end{list}

\para{RQ1: Model Validation}%
We assessed the model’s performance using previously \textbf{unseen data}. We sampled \sampleVal records from 5 different datasets
of 3 real-world software projects \cite{brownlee_2024_projects_dataset} 
\cite{DBLP:journals/ase/BrownleeCEGHPSS25}, none of which was used in~\cref{sec:modelextraction} to train or test the model.
We automatically constructed a dataset of \sampleVal (\diffP, \sumLLMshort{}) pairs. Then, we manually assigned them to the categories in Table \ref{tab:offline:res:15words} and afterwards independently tagged them with \model{}, 
resulting in the final format: (\diffP, auto-generated \sumLLMshort{}, manually tagged category, auto-tagged category). 
A cleanup stage was required before invoking \model{} to remove all quotation marks, backticks, the padded phrase \textit{``Here is a 15-word summary:"} or \textit{``Java diff:"}, and to adopt the verbs to fit the GI wordings (\eg{} change ``update" to ``modify").
To evaluate the quality of the model on unseen projects, we compared the manually and automatically assigned categories, 
out of which 75 did not match, while 143 were consistent.
Table~\ref{tab:acc_size} summarises per-category accuracy and overall accuracy. There are 6 categories to which no patch was manually assigned. 15 manual tags were invalid or inconclusive, suggesting a need to improve tagging, descriptions, or clustering granularity.

\noindent\textbf{Example 1: }%
A \sumLLMshort{} was \textit{``Java code change: catching ParseException with variable name ``e" and added try-catch block"} misdescribed a \diffP{} with no new try-catch block. The manual tag was left blank, while \model{} assigned Category~\#9.

\noindent\textbf{Example 2: }%
A \sumLLMshort{} was \textit{``Java code modified to handle timezone offset and check for daylight saving time"} matched no existing category and was arbitrarily assigned Category~\#4.\looseness=-1

\begin{figure}[t!]
  \centering
  \includegraphics[width=0.92\linewidth, trim=5pt 5pt 2pt 10pt, clip]{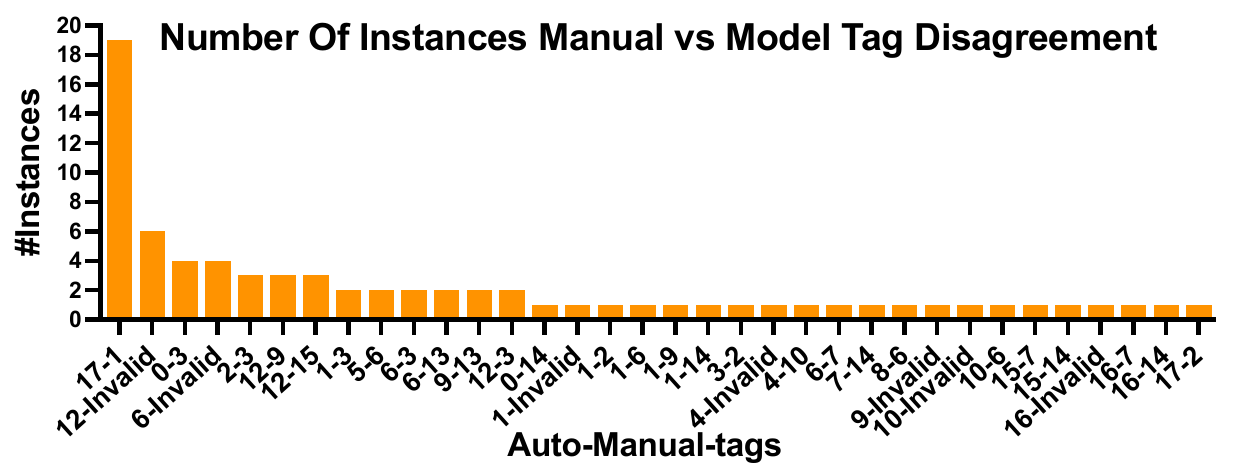}
  \caption{Mismatch Frequencies Manual Vs. Auto Patch.
  X-axis shows category pairs \texttt{a--b}: \texttt{a} is an automatically predicted category by the model \& \texttt{b} is the category assigned manually.}
  \label{fig:manual_auto_mismatch}
\end{figure}

\noindent
We analyzed the remaining 60 inconsistencies in Fig.~\ref{fig:manual_auto_mismatch}, which shows the differences between the manual and automated tags. Each x-axis label denotes an auto–manual pair (\eg{} 0–3 means auto-tagged as Category~\#0, manually as~\#3); the y-axis shows the count of such cases..
The largest mismatch involved 19 instances manually tagged as Category~\#1 (``No change") but \model{} assigned them to Category~\#17 (``Added dead code"), likely reflecting \model{}’s difficulty distinguishing edits with no behavioral effect.

We identified possible causes for the inconsistencies: 
\textbf{Missing categories}, as shown in Example 2; 
\textbf{Different prioritization} by \model{} and humans, \eg{} a case was tagged \#12 by \model{} and \#9 by a human who noted: 
\textit{``12 could also be considered but 9 is more important..."} for \sumLLMshort{}: \textit{``A Java code diff with 4 changes: catches ParseException, adds variable, and updates logic"};  and
\textbf{Incomplete or unclear summaries}, see example in \cite{anon_2025_15834984}.

\begin{tcolorbox}[colback=gray!20, colframe=black,boxrule=0.7pt, before skip=5pt, after skip=5pt, left=2pt,
  right=2pt,
  top=2pt,
  bottom=2pt]
    \textbf{RQ1 Answer. }
On unseen data from five datasets, \model{} achieved 66\% accuracy, lower than on test data in~\cref{sec:modelextraction}. 
For generalization, there is a need to  
cleaning LLM-generated text before model input,  
adapting new categories for new projects, and  
adding more manually tagged data to reduce misprioritisation, to enhance \model's reliability for decision-making in \gin{}.
\end{tcolorbox}

\para{RQ2: Patches Quality}%
We used \model{} on a dataset of 3,232 patches from~\cite{DBLP:conf/ssbse/BrownleeCEGHPSS23,DBLP:journals/ase/BrownleeCEGHPSS25}, 
for 5 projects
and generated by 3 LLM models \cite{brownlee_2024_projects_dataset}.
Further, we report the number of patches that \textit{compile}, \textit{passed} all tests or are \textit{NoOps} in each category.

\begin{figure}[t!]
  \centering
  \includegraphics[width=0.89\linewidth, trim=5pt 20pt 2pt 2pt, clip]{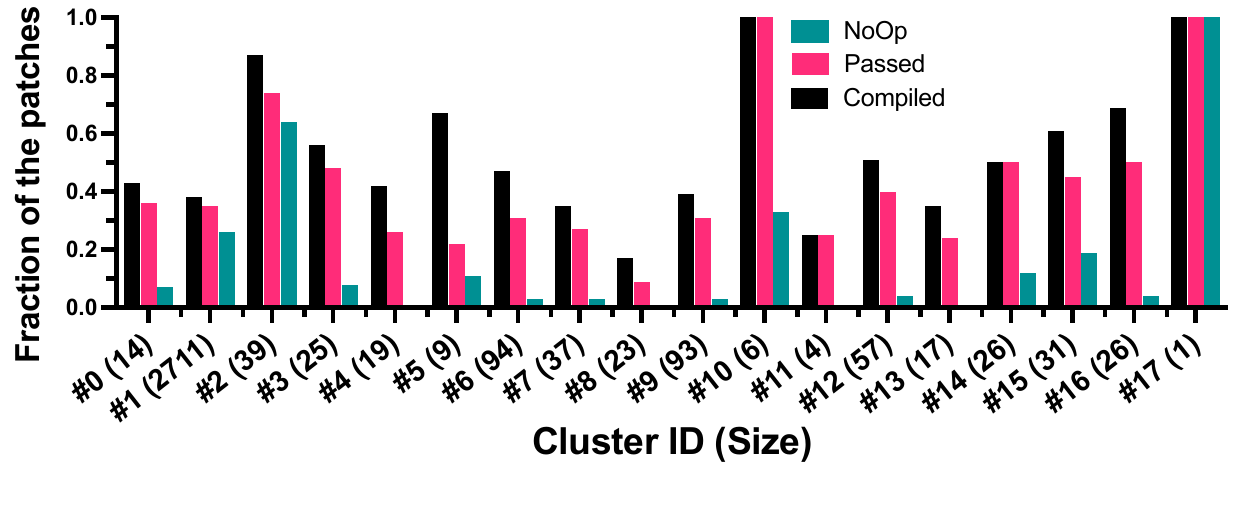}
  \caption{Patch Categorizing Results.
  X-axis label per Category ID (its size) \& Y-axis is the fraction of patches compiled, passed all unit tests and did not perform any operation (NoOp).
  }
  \label{fig:offline:res:stat}
\end{figure}

Fig.~\ref{fig:offline:res:stat} summarizes the patch statistics of the automatically assigned patches. \textit{Compiled} and \textit{Passed} report the number of patches per category that compiled, and passed all tests, respectively. \textit{NoOp} indicates the proportion of semantically neutral patches, \ie{} those with no observable effect.

The most frequent category is ``No change" (Category 1, 2,711 patches), 
followed by much smaller categories:
94 patches are in ``Changes to method names" (Category 6), 
93 in ``Added exception-handling constructs (unreachable or reachable)" (Category 9), and
57 in ``Modified Variable/Class/Object Name" (Category 12). 
Except for Categories 1–2 and 17, there should be no NoOps. 
Manual inspection of misclassified cases revealed misleading \sumLLMshort{}, \eg{} \textit{“Special case added to look for partial maps within a list in code”}, despite the switch statement remaining unchanged. 
Categories 3–16, likely with observable effects, have compilation rates from 0.17 to 1.00.
\begin{table}[t!]
\centering
\vspace{-0.2 cm}

\caption{Per-category Clustering Accuracy \& Category Size.}
\setlength{\tabcolsep}{4pt}
\footnotesize
\begin{tabular}{rccccccccc}
\toprule
\textbf{Cat.} & 0   & 1    & 2    & 3    & 4  & 5  & 6    & 7    & 8    \\
\textbf{Acc.} & NA  & 0.85 & 0.50 & 0.00 & NA & NA & 0.00 & 0.00 & 1.00 \\
\textbf{Size} & 0   & 129  & 6    & 13   & 0  & 0  & 5    & 3    & 1    \\
\midrule
\textbf{Cat.} & 9    & 10   & 11 & 12   & 13   & 14   & 15   & 16 & 17 \\
\textbf{Acc.} & 0.78 & 0.50 & NA & 1.00 & 0.20 & 0.55 & 0.40 & NA & NA \\
\textbf{Size} & 18   & 2    & 0  & 5    & 5    & 11   & 5    & 0  & 0   \\
\bottomrule
\multicolumn{10}{r}{\textbf{Overall accuracy:} 0.66 with 218 manually tagged instances.} \\
\end{tabular}
\label{tab:acc_size}
\end{table}
The lowest test passing rate is 0.09 (Category 8, 0.17 compilation rate). The sharpest relative drop appears in Category~5, where only 0.22 of the patches passed all tests despite a 0.67 compilation rate, \ie{} just one-third of the compiled patches were successful. The high rate of NoOp failures is attributed to this class including approximated 37\% of suggested patches that included no code, so no diff, and were counted as fails in \cite{brownlee_2024_projects_dataset}.
In a GI context, one would prefer patches in categories like 7 and 16, where the fraction of passing tests is high but the number of NoOps is low: these are patches that change the code implementation in some way that preserves functionality (at least with respect to the tests). This can form the basis of a filtering scheme as we propose in our research agenda (\cref{researchAgenda}).

\begin{tcolorbox}[colback=gray!20, colframe=black,boxrule=0.7pt, before skip=5pt, after skip=5pt, left=2pt,
  right=2pt,
  top=2pt,
  bottom=2pt]
    \textbf{RQ2 Answer. }
    The high rate of NoOp patches suggests that filtering them using \model{}, a lightweight alternative to LLM calls, can improve \gin{}’s performance. Moreover, patches that compile and have an observable effect tend to have a good chance of passing the test suite, making them valuable in the context of GI.
\end{tcolorbox}

\section{Conclusion \& Research Agenda}
\label{researchAgenda}
We introduced \model{} for classifying automatically generated software patches and to gain deeper insights into the nature of LLM-based edits. \model{} classified patches into 18 different categories and achieved 66\% accuracy for patches generated for five unseen real-world software projects. Moreover, \model{} enabled us to detect \textit{NoOp} edits in advance and, prospectively, skip the execution of the test suite in many situations to save resources.

We envisage future work to leverage the insights from our work to further improve \model{}, for example, one can train \model{} with patches generated for additional software projects to have a broader and more diverse data base so that the classifier is optimally equipped for practical usage. Further, we will study the usage of LLMs to improve the classifier \cite{miller2025human}.  
In addition to LLM-based edits, the improved classifier can be used to analyze traditional mutation methods and compare the results with the LLM-generated patches in order to learn when a specific edit type is more useful or when a combination is more beneficial. Our expectation is that while LLM-generated patches are often semantically more adequate,  traditional mutations introduce more diversity \cite{tevc2024comparison}, which can lead to solutions that would not have been possible with LLMs alone. A combination could unlock the advantages of both methods for GI and can be efficiently obtained by using appropriate heuristics and ML methods \cite{EAPR2021}.
Furthermore, \model{} can be used to improve the search for patches with \gin{}.
To this end, one can implement Fig. ~\ref{fig:online} - step \circled{J} 
which classifies the patches generated by \gin{} during the run.
Based on the assigned class, it can be decided whether an evaluation with the test suite should be performed. For patch classes that have shown a high success rate so far, it is worth running the test suite. For classes with a low success rate, \eg{} the evaluation can be skipped for \textit{modified a comment} or \textit{added dead code}. 
Given \model{}\textit{'s} ability to detect \textit{NoOp} edits in advance, we expect to save a significant amount of compute for the evaluation in \gin{}.
Additionally, one can provide the user with helpful information in addition to the suggested patches. Last but not least, explanations for the respective patches \cite{ssbse2023evaluating} can be automatically generated to describe what led to performance improvements.

\bibliographystyle{IEEEtran}
\bibliography{main}

\end{document}